\documentclass[prd,preprint,showpacs]{revtex4}
\usepackage{amsmath}
\usepackage{latexsym}
\usepackage{amsfonts}
\usepackage{graphicx}
\begin{document}

\title{Constraints on Cardassian universe from Gamma ray bursts}
\author{Tai-Shan Wang}
\affiliation{Department of Physics \& Center for Astrophysics,
Tsinghua University, Beijing 100084, China}
\author{Nan~Liang\,}
\email[\,email address:\ ]{liangn@bnu.edu.cn} \affiliation{
Department of Astronomy, Beijing Normal University, Beijing 100875,
China}

\begin{abstract}
Constraints on the original Cardassian model and the modified
polytropic Cardassian model are examined from the recently derived
42 gamma-ray bursts (GRBs) data  calibrated with the method avoiding
the circularity problem. The results show that GRBs can be an
optional observation to constrain on the Cardassian models.
Combining the GRBs data with the newly derived size of baryonic
acoustic oscillation peak from the Sloan Digital Sky Survey (SDSS),
and the position of first acoustic peak of the Cosmic Microwave
Background radiation (CMB) from Wilkinson Microwave Anisotropy Probe
(WMAP), we find  $\Omega_{m0}=0.27_{-0.02}^{+0.02},
n=0.06_{-0.08}^{+0.07}$ ($1\sigma$) for the original Cardassian
model, and $\Omega_{m0}=0.27^{+0.03}_{-0.02}$,
$n=-0.09_{-1.91}^{+0.23}, \beta=0.82_{-0.62}^{+2.10}$ ($1\sigma$)
for the modified polytropic Cardassian model.
\end{abstract}

  \maketitle

\section{Introduction}
The astrophysical observations of recent years, including Type Ia
supernovae (SNe Ia;
~\cite{bie98,bie04,per99,ton03,kno03,bar04,kow08}),the size of the
baryonic acoustic oscillation (BAO) peak detected in the large-scale
correlation functions of luminous red galaxies from the Sloan
Digital Sky Survey (SDSS; \cite{eis05}), and cosmic microwave
background radiation (CMB;
\cite{bal00,ber00,jaf01,spe03,spe07,komtemp}), support that the
present expansion of our universe is accelerating. This is important
to help us understand our universe, but its nature still remains as
an open question today. A large number of cosmological models have
been proposed by cosmologists, in order to explain the accelerating
expansion of the universe. There are two main categories of
proposals. The first ones (dark energy models) are proposed by
assuming an energy component with negative pressure (the dark
energy) in the universe, this dark energy dominates the total energy
density of the universe and drives its acceleration of expansion at
recent times.
The other proposals suggest that the general relativity fails in the
present cosmic time space scale, and the extra geometric effect is
the reason of the acceleration. 
The Cardassian models which investigate the acceleration of the
universe by a modification to the Friedmann-Robertson-Walker (FRW)
equation \cite{fre02} belongs to this categorie.

 Here we focus on
the Cardassian models, including both of the original Cardassian
model and the modified polytropic Cardassian model. The original
Cardassian model is based on the modified Friedmann equation and has
two parameters $\Omega_{m0}$ and $n$. It predicts the same
distance-redshift relation as generic quintessence models, although
their physical principles are totally different from each other. The
modified polytropic Cardassian model can be obtained by introducing
an additional parameter $\beta$ into the original Cardassian model
which reduces to the original model when $\beta=1$. The luminosity
distance-redshift relation of the modified polytropic Cardassian
model can be very different from generic quintessence models.

As we know, many observational constraints have been placed on
Cardassian models, including those from the angular size of high
redshift compact radio sources \cite{zhu02}, the distance modula of
SNe Ia
\cite{wan03,dav07,zhu03,cao03,szy04,god04,fri04,zhu04,ben05,ama05,ben06},
the shift parameter of the CMB\cite{dav07,fri04,ama05,yi07}, the
baryon acoustic peak from the SDSS \cite{dav07,yi07}, the
gravitational lensing \cite{alc05}, the x-ray gas mass fraction of
clusters \cite{zhu04,zhu042}, the large scale structure
\cite{ama05,mul03,fay06}, the Hubble parameter versus redshift data
\cite{yi07}, and the combined analysis of different data
\cite{wan07}.

Recently, the gamma-ray bursts (GRBs) have been regarded as the
standard candles since several empirical GRB luminosity relations
were proposed as distance indicators to be a complementary probe to
the universe \cite{ghi06,zha07,mes06,woo06,sch07}. However, an
important point related to the use of GRBs for cosmology is the
dependence on the cosmological model in the calibration of GRB
relations. The relations of GRBs presented above have been
calibrated by assuming a particular cosmological model for the
difficulty to calibrate the relations with a low-redshift sample.
Therefore the circularity problem can not be avoided
easily\cite{ghi06}. A new method in a completely cosmology
independent manner to calibrate GRBs by interpolating directly from
SNe Ia has been proposed\cite{lia08,lia08b}, and the circularity
problem can be solved. Following the SNe Ia interpolation method,
the distance modulus of 42 calibrated GRBs at $z>1.4$ are derived.
Now, one may use them to constrain cosmological models without
circularity problem\cite{wei08}.

The main purpose of this work is to give out constraints on
Cardassian models with the newly derived 42 GRBs' data, which have
avoided the circularity problem by new method \cite{lia08,lia08b},
along with the size of baryonic acoustic oscillation peak from SDSS
\cite{eis05}, and the position of first acoustic peak of CMB from
WMAP \cite{komtemp}. As a result, we find that stronger constraints
can be given out with this combined data set than most of the former
results, such as the results with SNe Ia
data\cite{szy04,god04,fri04}, and the results with other combined
data set\cite{zhu04,wan07}.

This paper is organized as follows: In section 2, we give out the
basic equations of Cardassian models. In section 3, we describe the
analysis method for the GRBs data and present the constraint
results. In section 4, we describe the analysis method for the
combined data set including GRBs, BAO and CMB, and present their
constraint results. In section 5, we give out the conclusions and
some discussions.

\section{THE BASIC EQUATIONS OF Cardassian MODELS}
In 2002, Freese and Lewis \cite{fre02} proposed Cardassian model as
a possible explanation for the acceleration by modifying the FRW
equation without introducing the dark energy. The basic FRW equation
can be written as
\begin{equation}
H^2=\frac{8\pi G}{3}\rho
\end{equation}
where G is the Newton gravitation constant and $\rho$ is the density
of summation of both matter and vacuum energy. For the Cardassian
model, which is modified by adding a term on the right side of
Eq.(1), the FRW equation is shown as below
\begin{equation}
H^2=\frac{8\pi G}{3}\rho_m+B\rho_m^n
\end{equation}
The latter term, which is so called Cardassian term, may show that
our observable universe as a $3+1$ dimensional brane is embedded in
extra dimensions. Here $n$ is assumed to satisfy $n<2/3$, and
$\rho_m$ only represents the matter term without considering the
radiation for simplification. The first term in Eq.(2) dominates
initially, so the equation becomes to the usual Friedmann equation
in the early history of the universe. At a redshift $ z\sim O(1)$
\cite{fre02}, the two terms on the right side of the equation become
equal, and thereafter the second term begins to dominate, and drives
the universe to accelerate. If $B=0$, it becomes the usual FRW
equation, but with only the density of matter. If $n=0$, it is the
same as the cosmological constant universe. By using

\begin{equation}
\rho_m=\rho_{m0}(1+z)^3=\Omega_{m0}\rho_c(1+z)^3
\end{equation}
we obtain

\begin{equation}
E^2=\frac{H^2}{H_{0}^2}=\Omega_{m0}(1+z)^3+(1-\Omega_{m0})(1+z)^{3n}
\end{equation}
where $z$ is the redshift, $\rho_{m0}$ is the present value of
$\rho_m$ and $\rho_c=3H_0^2/8\pi G$ represents the present  critical
density of the universe. Obviously, this model predicts the same
distance-redshift relation as the quiessence with $\omega_Q=n-1$,
but with totally different intrinsic nature.

The luminosity distance of this model is
\begin{equation}
d_L=cH_0^{-1}(1+z)\int_0^zdz[\Omega_{m0}(1+z)^3+(1-\Omega_{m0})(1+z)^{3n}]^{-1/2}
\end{equation}
where $c$ is the velocity of light.

The modified polytropic Cardassian universe is obtained by
introducing an additional parameter $\beta$ into the original
Cardassian model, which reduces to the original model if $\beta=1$,

\begin{equation}
H^2=H_{0}^2[\Omega_{m0}(1+z)^3+(1-\Omega_{m0})f_X(z)]
\end{equation}
where

\begin{equation}
f_X(z)=\frac{\Omega_{m0}}{1-\Omega_{m0}}(1+z)^3[(1+\frac{\Omega_{m0}^{-\beta}-1}{(1+z)^{3(1-n)\beta}})^{1/\beta}-1]
\end{equation}
Here if the $\beta=1$ and $n=1$, then $f_X(z)=1$ , and this model
just corresponds to $\Lambda$CDM.  The corresponding luminosity
distance of Eq. (6) is
\begin{equation}
d_L=cH_0^{-1}(1+z)\int_0^zdz[\Omega_{m0}(1+z)^3[1+\frac{\Omega_{m0}^{-\beta}-1}{(1+z)^{3(1-n)\beta}}]^{1/\beta}]^{-1/2}
\end{equation}

\section{CONSTRAINTS FROM GRBs}
The distance modulus of the 42 GRBs ($z>1.4$) we use here are newly
obtained by the interpolating method\cite{lia08} which compiled in
Table 2 of ref. \cite{lia08b}. The main idea is the cosmic distance
ladder. Similar to the case of calibrating SNe Ia as the standard
candles by using Cepheid variables, we can also calibrate GRBs as
standard candles with a large amount of SNe Ia. These distance
modulus are so far the most independent GRBs' result on prior
cosmological models, and their method avoids the circularity problem
more clearly than previous cosmology-dependent calibration methods.
Although the number of GRBs is small and the systematic and
statistical errors are relatively large so that their contribution
to the constraints would be not so significant, this is still a
beneficial exploration. So far, this data set has never been used to
constrain the Cardassian models, here for the first time, we
introduce them into the constraining process. Constraints from GRBs
can be obtained by fitting the distance modulus $\mu(z)$
\begin{equation}
\mu(z)=5\log_{10}d^L+M
\end{equation}
Here $M$ being the absolute magnitude of the object, which is 42.38,
and we set $H_0=72\ km s^{-1}Mpc^{-1}$~\cite{Freedman2001}.

In order to place limits on model parameters ($\Omega_{m0}, n,
\beta$) with the observation data, we make use of the maximum
likelihood method, that is, the best fit values for these parameters
can be determined by minimizing
\begin{equation}
\chi^2_{GRBs}=\sum_i
\frac{[\mu_{obs}(z_i)-\mu(z_i)]^2}{\sigma_i^2}\;,
\end{equation}
where the $\sigma_i$ represent the uncertainty of GRBs data.

With the 42 GRBs data set, by minimizing the corresponding
$\chi^2_{GRBs}$ in Eq. (10), we get the constraints results as Fig.1
shows. This result is consistent with the former result of SNe Ia
small sample\cite{cao03}.  Fig.2 shows the result for the modified
polytropic Cardassian model with the GRBs data set only.

\section{CONSTRAINTS FROM COMBINING GRBs, BAO AND CMB}
In 2005, Eisenstein et al.\cite{eis05} successfully found the size
of baryonic acoustic oscillation peak by using a large spectroscopic
sample of luminous red galaxy from the SDSS and obtained a parameter
$A$, which is independent of dark energy models and for a flat
universe can be expressed as
\begin{equation}
A=\frac{\sqrt{\Omega_{m0}}}{E(z_1)^{1/3}}[\frac{1}{z_1}\int_0^{z_1}\frac{dz}{E(z)}]^{2/3}
\end{equation}
where $z_{1}=0.35$ and the corresponding $A$ is measured to be $A =
0.469\pm0.017$. Using parameter $A$ we can obtain the constraint on
Cardassian models from the SDSS.

The shift parameter $R$ of the CMB data can be used to constrain the
Cardassian models and it can be expressed as \cite{bon97}
\begin{equation}
R=\sqrt{\Omega_{m0}}\int_0^{z_r}\frac{dz}{E(z)}
\end{equation}
here $z_r=1089$ for a flat universe. From the five-year WMAP result
\cite{spetemp}, the shift parameter is constrained to be
$R=1.700\pm0.019$ \cite{komtemp}.

The best fit values for model parameters can be determined by
minimizing
\begin{equation}
\chi^2=\sum_i
\frac{[\mu_{obs}(z_i)-\mu(z_i)]^2}{\sigma_i^2}+\frac{(A-0.0469)^2}{0.017^2}+\frac{(R-1.700)^2}{0.019^2}\;.
\end{equation}

With the GRBs + BAO + CMB data set, we find
$\Omega_{m0}=0.27_{-0.02}^{+0.02}, n=0.06_{-0.08}^{+0.07}$ for the
original Cardassian model at $1\sigma$ confidence level. Details for
constraints are shown in Fig. 3. We find that combining these
observational data can tighten these model parameters significantly
comparing to the results from former academic papers
\cite{gon03,wan03,sav05,yi07}.

For the modified polytropic Cardassian model,   we obtain
$\Omega_{m0}=0.27^{+0.03}_{-0.02}$, $n=-0.09_{-1.91}^{+0.23},
\beta=0.82_{-0.62}^{+2.10}$ at the $1 \sigma$ confidence level.
Details for constraints are shown in Fig. 4.

From the Figs. 3 and 4 we find  the flat $\Lambda$CDM cosmology is
consistent with observations since the  Cardassian model reduces to
the flat $\Lambda$CDM when $n=0$ and the modified polytropic
Cardassian model is  $\beta=1, n=0$.

\section{Conclusions}
From our data analysis results (Fig. 1 and Fig. 2), we can conclude
that the 42 newly derived GRBs data can be used to set constraints
on the Cardassian models. On the other hand, with GRBs + BAO + CMB
jointly analysis, we obtain $\Omega_{m0}=0.27_{-0.02}^{+0.02},
n=0.06_{-0.08}^{+0.07}$ for the original Cardassian model, and
$\Omega_{m0}=0.27^{+0.03}_{-0.02}$, $n=-0.09_{-1.91}^{+0.23},
\beta=0.82_{-0.62}^{+2.10}$ for the modified polytropic Cardassian
model at $1\sigma$ confidence level.

It is worth noticing that GRBs are important potential probes for
cosmic history up to $z>6$. Due to the lack of enough low red-shift
GRBs to calibrate the luminosity relation, GRBs can not be used
reliably and extensively in cosmology for now, but ref. \cite{lia08}
has made an important improvement to this. Hereafter, along with
more observed GRBs, like these 42 GRBs data, whose distance modulus
are calibrated with the method excluding the circularity problem,
GRBs could be used as an optional choice to set tighter constraints
on parameters of Cardassian models and even other cosmographic model
parameters.

Recently, some authors point out that there is observational
selection bias in GRB relations\cite{But07,Sha09} and possible
evolution effects in GRB relations have been dicussed
\cite{Li07,Tsu08}. However, Ghirlanda et al. \cite{Ghi08} found that
no sign of evolution with redshift of one GRB relation (the Amati
relation), and the instrumental selection effects do not dominate
for GRBs detected before the launch of the Swift satellite. More
recently, ref. \cite{Ghi09} indicate that another GRB relation (the
$E_p-L_{iso}$ relation) is not the result of instrumental selection
effects. Nevertheless, for considering GRBs as standard candles for
cosmological use, further examinations of possible evolution effects
and selection bias should be required.

\begin{figure}
\begin{center}
\includegraphics[angle=0,scale=0.5]{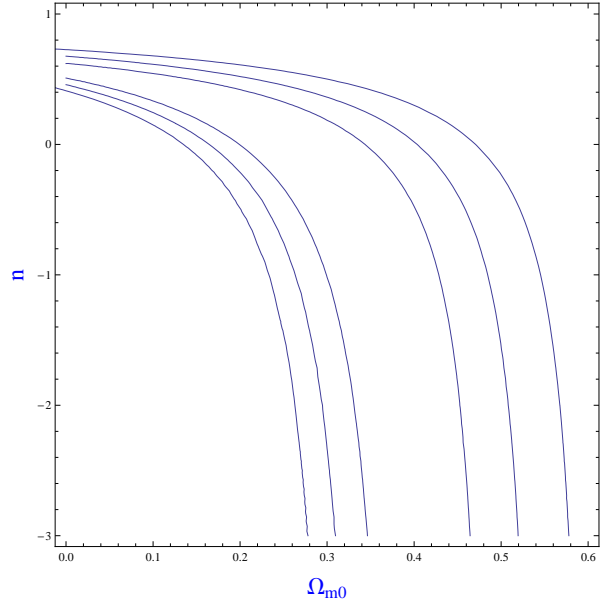}
\end{center}
\caption{Constraints on $\Omega_{m0}$ and $n$ from $1\sigma$ to
$3\sigma$ are obtained from 42 GRBs data for the original Cardassian
model. }
\end{figure}

\begin{figure}
\begin{center}
\includegraphics[angle=0,scale=0.5]{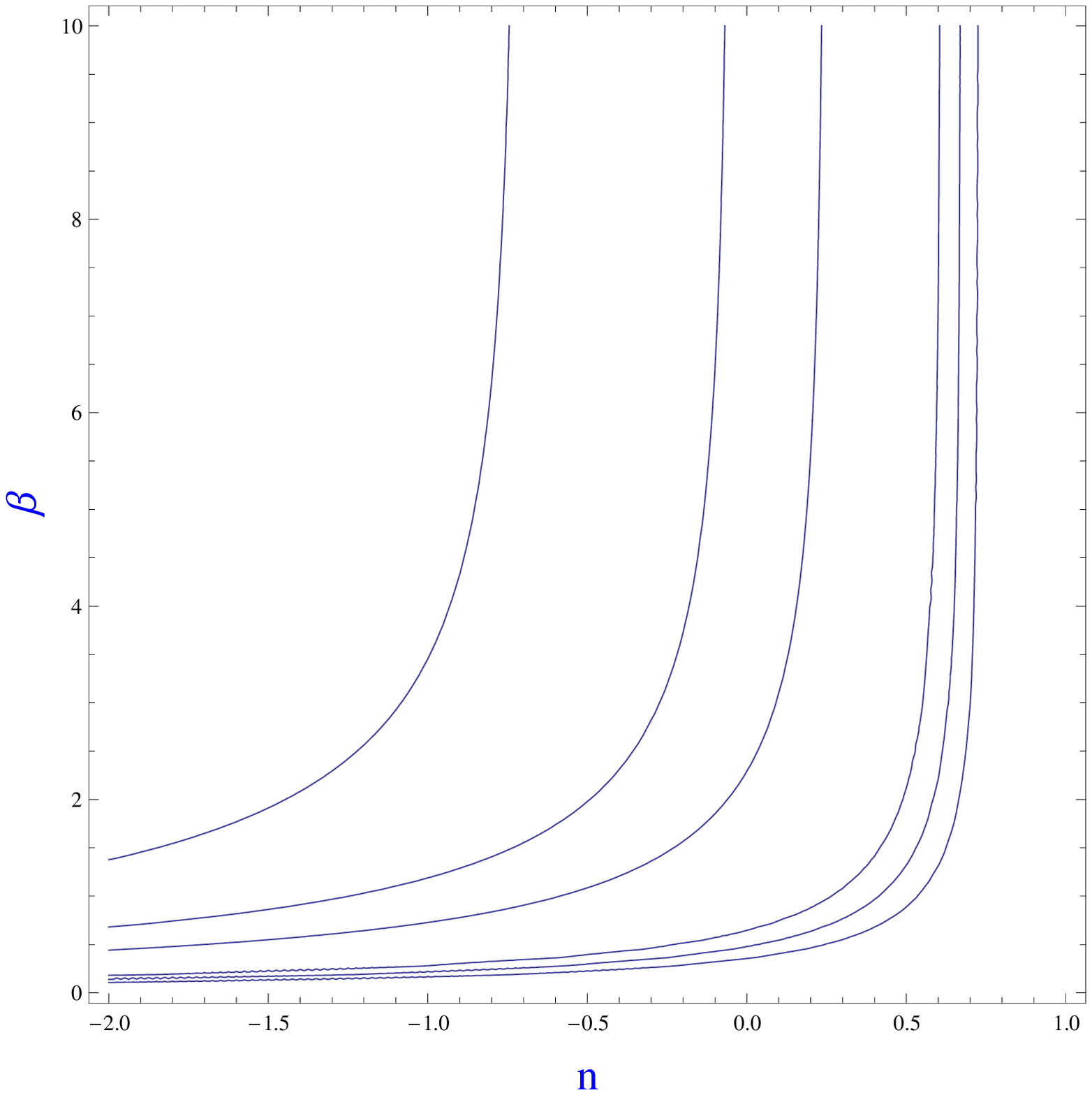}
\includegraphics[angle=0,scale=0.5]{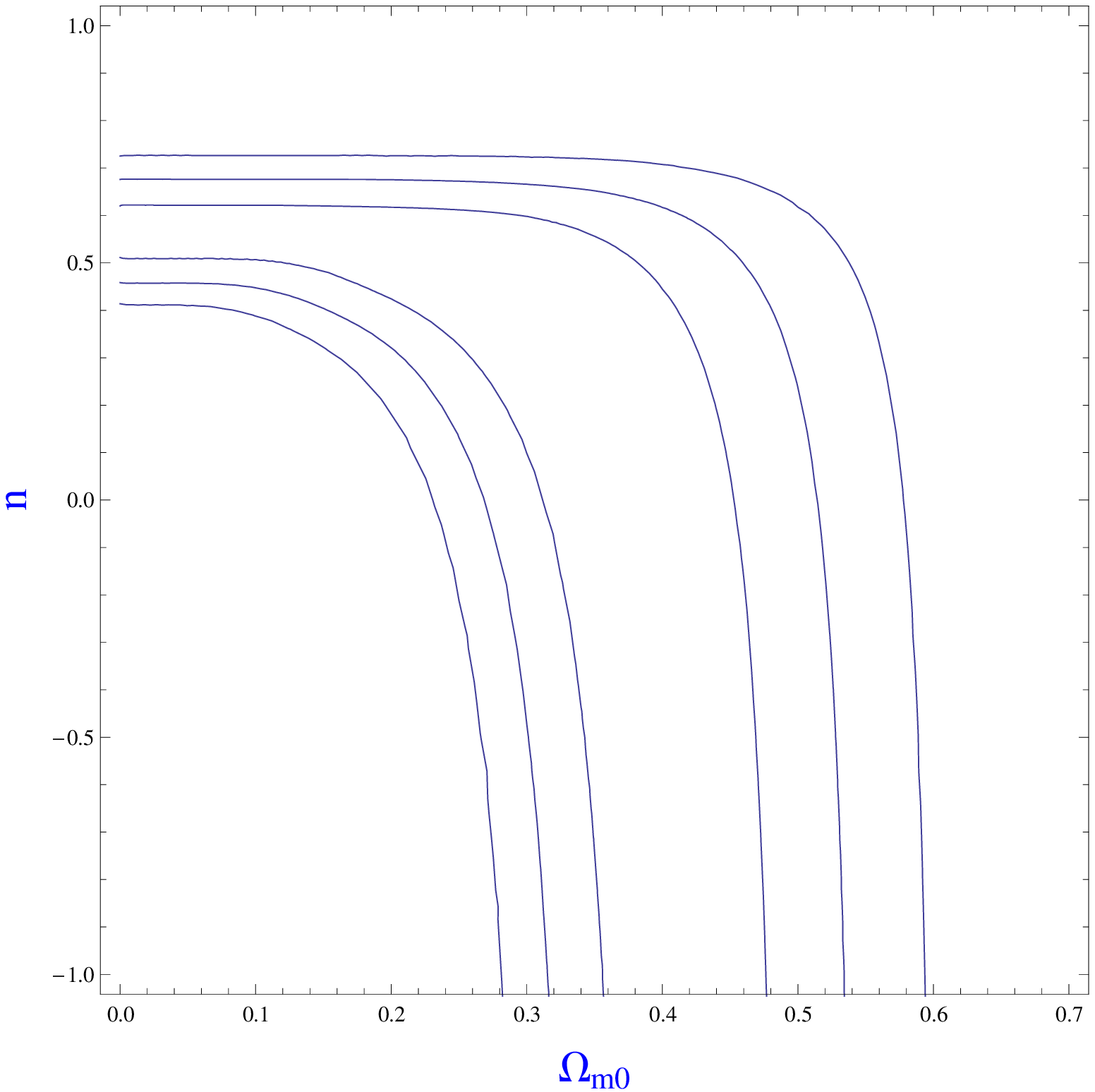}
\end{center}
\caption{Constraints on parameters of the modified polytropic
Cardassian model by setting the best fit value over
$\Omega_{m0},\beta$ respectively from $1\sigma$ to $3\sigma$ are
obtained from 42 GRBs data.}
\end{figure}

\begin{figure}
\begin{center}
\includegraphics[angle=0,scale=0.5]{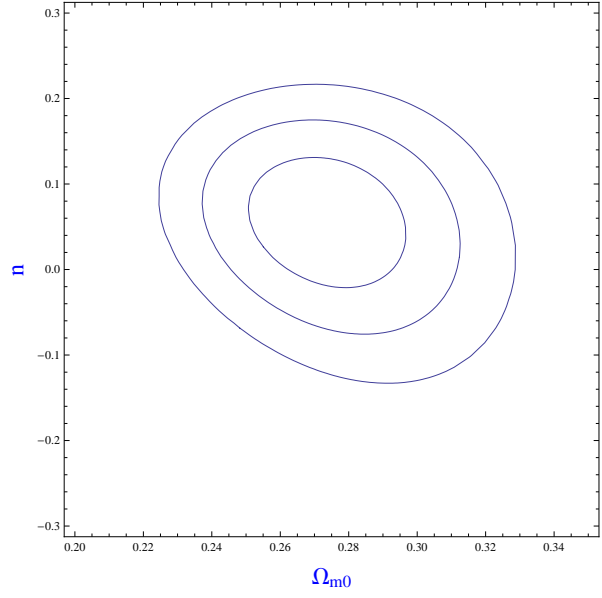}
\end{center}
\caption{Constraints on $\Omega_{m0}$ and $n$ from $1\sigma$ to
$3\sigma$ are obtained from combined data set (including
GRBs+BAO+CMB) for the original Cardassian model.}
\end{figure}

\begin{figure}
\begin{center}
\includegraphics[angle=0,scale=0.5]{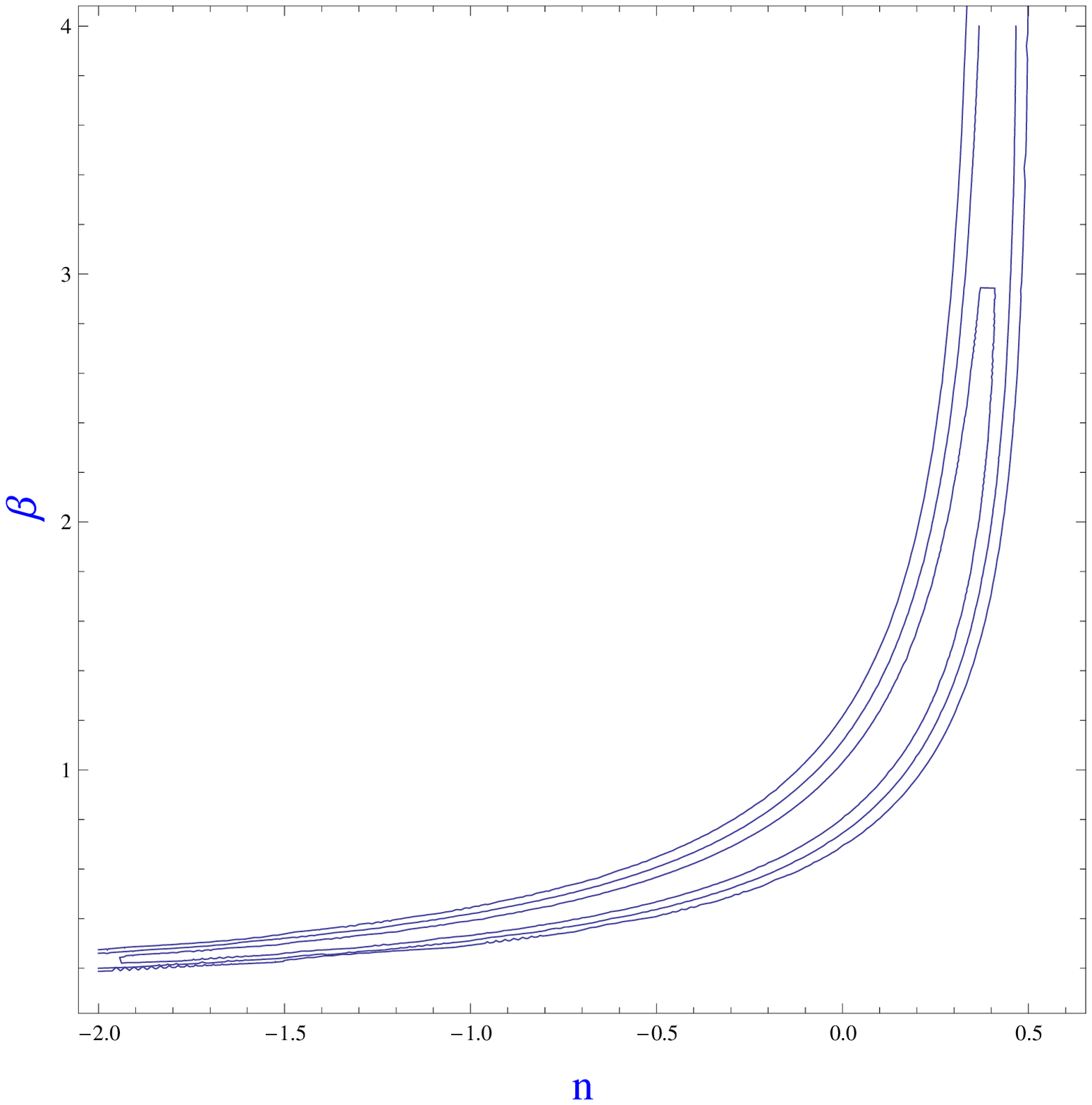}
\includegraphics[angle=0,scale=0.5]{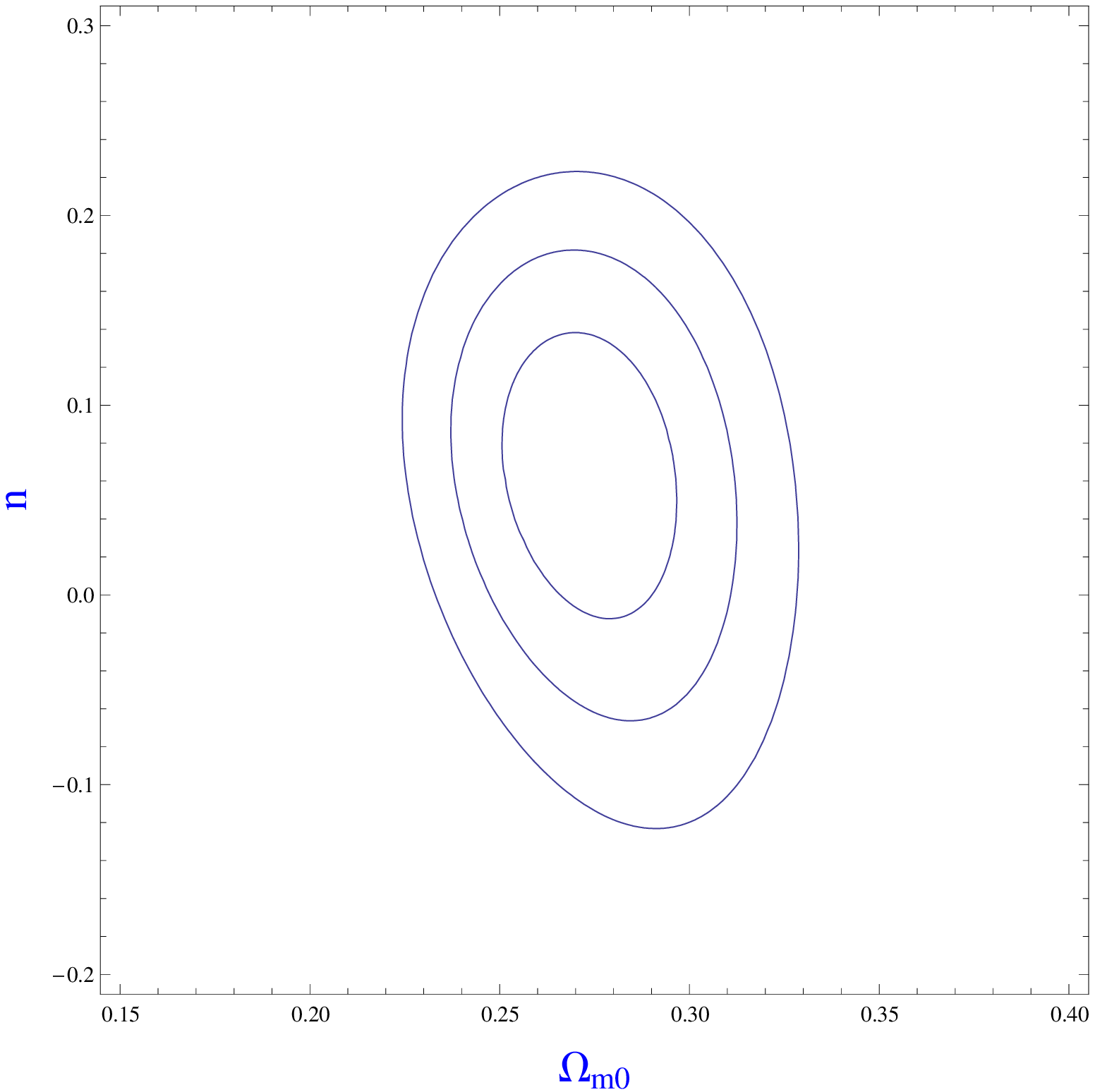}
\end{center}
\caption{Constraints on parameters of the modified polytropic
Cardassian model by setting the best fit value over
$\Omega_{m0},\beta$ respectively from $1\sigma$ to $3\sigma$ are
obtained from combined data set.}
\end{figure}



\begin{thebibliography}{sp99}
\bibitem{bie98} Riess A. G., et al. 1998, AJ 116, 1009
\bibitem{bie04} Riess A. G., et al. 2004, ApJ 607, 665
\bibitem{per99} Perlmutter S., et al. 1999, ApJ 517, 565
\bibitem{ton03} Tonry J. L., et al. 2003, ApJ 594, 1
\bibitem{kno03} Knop R. A., et al. 2003, ApJ 598, 102
\bibitem{bar04} Barris B. J., et al. 2004, ApJ 602, 571
\bibitem{kow08} Kowalski M., et al. 2008, Astrophys. J. 686, 749
\bibitem{eis05} Eisenstein D. J., et al. 2005, ApJ 633, 560
\bibitem{bal00} Balbi A., et al. 2000, ApJ 545, L1
\bibitem{ber00} de Bernardis P., et al. 2000, Nature 404, 955
\bibitem{jaf01} Jaffe A. H., et al. 2001, Phys. Rev. Lett. 86, 3475
\bibitem{spe03} Spergel D. N., et al. 2003, ApJS 148, 175
\bibitem{spe07} Spergel D. N., et al. 2007, ApJS 170, 377S
\bibitem{komtemp} Komatsu E., et al. 2009, ApJS 180, 330
\bibitem{fre02} Freese K., \& Lewis M. 2002, Phys. Lett. B 540, 1
\bibitem{zhu02} Zhu Z. H., \& Fujimoto M. K. 2002, Astrophys. J. 581, 1
\bibitem{wan03} Wang Y., Freese K., Gondolo P., \& Lewis M. 2003, Astrophys. J. 594, 25
\bibitem{dav07} Davis T. M., et al. 2007, ApJ 666, 716D
\bibitem{zhu03} Zhu Z. H., \& Fujimoto M. K. 2003, Astrophys. J. 585, 52
\bibitem{cao03} Cao Li, 2003, Chin. J. Astron. Astrophys. Vol. 3 No. 4, 341¨C346
\bibitem{szy04} Szydlowski M., \& Czaja W. 2004, Phys. Rev. D 69, 083507
\bibitem{god04} Godlowski W., Szydlowski M., \& Krawiec A. 2004, Astrophys. J. 605, 599
\bibitem{fri04} Frith W. J. 2004, Mon. Not. R. Astron. Soc. 348, 916D920
\bibitem{zhu04} Zhu Z. H., Fujimoto M. K., \& He X. T. 2004, Astrophys. J. 603, 365
\bibitem{ben05} Bento M. C., Bertolami O., Santos N. M. C., \& Sen A. A. 2005, Phys. Rev. D 71, 063501
\bibitem{ben06} Bento M. C., Bertolami O., Reboucas M. J., \& Santos N. M. C. 2006, Phys. Rev. D 73, 103521
\bibitem{ama05} Amarzguioui M., Elgaroy O., \& Multamaki T. 2005, JCAP 01, 008
\bibitem{yi07} Yi Z. L., \& Zhang T. J. 2007, Phys. Rev. D 75, 083515
\bibitem{alc05} Alcaniz J. S., Dev A., \& Jain D. 2005, Astrophys. J. 627, 26 
\bibitem{zhu042} Zhu Z. H., \& Fujimoto M. K. 2004, Astrophys. J. 602, 12
\bibitem{mul03} Multamaki T., Gaztanaga E., \& Manera M. 2003, Mon. Not. R. Astron. Soc. 344, 761
\bibitem{fay06} Fay S., \& Amarzguioui M. 2006, Astron. Astrophys. 460, 37
\bibitem{wan07} Wang F. Y. 2007, JCAP 08, 020
\bibitem{ghi06} Ghirlanda G., Ghisellini G. \& Firmani C. 2006, New J. Phys. 8, 123
\bibitem{zha07} Zhang B. 2007, Chin. J. Astron. Astrophys. 7, 1
\bibitem{mes06} Meszaros P. 2006, Rept. Prog. Phys. 69, 2259
\bibitem{woo06} Woosley S. E., \& Bloom J. S. 2006, Ann. Rev. Astron. Astrophys. 44, 507
\bibitem{sch07} Schaefer B. E. 2007, Astrophys. J. 660, 16
\bibitem{lia08} Liang N., et al. 2008, Astrophys. J. 685, 354
\bibitem{lia08b} Liang N, Zhang S N, AIP Conf. Proc., 2008, 1065, 367
\bibitem{wei08} Wei H. 2008, EPJC, 2009, 60,449
\bibitem{Freedman2001}Freedman  W. L. et al., Astrophys. J. 553, 47 (2001).
\bibitem{bon97} Bond J. R., Efstathiou G., \& Tegmark M. 1997, Mon. Not. R. Astron. Soc. 291, L33.
\bibitem{spetemp} Spergel D. N., et al. 2007, ApJS 170, 377.  
\bibitem{gon03} Gondolo P., \& Freese K. 2003, Phys. Rev. D 68, 063509
\bibitem{sav05} Savage C., Sugiyama N., \& Freese K. 2005, JCAP 10, 007

\bibitem{But07} Butler, N R, et al.  ApJ, 2007, 671, 656
\bibitem{Sha09} Shahmoradi A,Nemiro R J.  arXiv:0904.1464
\bibitem{Li07} Li, L X. MNRAS, 2007, 379, L55
\bibitem{Tsu08} Tsutsui, R, et al. MNRAS, 2008, 386, L33
\bibitem{Ghi08} Ghirlanda G et al. MNRAS, 2008, 387, 319
\bibitem{Ghi09} Ghirlanda G et al. arXiv:0908.2807


\end{thebibliography}
\end{document}